\documentstyle[floats,multicol,aps,epsf,prb]{revtex}

\begin{document}
\draft

\twocolumn[\hsize\textwidth\columnwidth\hsize\csname @twocolumnfalse\endcsname

\title{Barriers in the p-spin interacting spin-glass model. The dynamical approach.}
\author{A.V. Lopatin and L. B. Ioffe}
\address{ Department of Physics, Rutgers University,
Piscataway, New Jersey 08855}
\date{\today}
\maketitle

\begin{abstract}

We investigate the barriers separating metastable states in the spherical
p-spin glass model using the instanton method. We show that the problem of
finding the barrier heights can be reduced to the causal two-real-replica
dynamics. We find the probability for the system to escape one of the
highest energy metastable states and the energy barrier corresponding to
this process.

\end{abstract}

 \vskip2pc]

One of the most important properties of the spin glasses is the existence of
large barriers separating metastable states. The calculation of the barrier
heights is not a simple problem even for the long range spin-glass models
for which the mean field theory is valid. Although the standard replica
approach \cite{Mezard87} allows one to find all thermodynamical properties,
these properties do not contain information about the barrier heights. The
usual dynamical mean field approach cannot be used either because it
involves taking the limit $N$ $\rightarrow \infty $ at an early stage of the
calculation, in this limit all barriers become infinite. The modification of
the replica approach allowing to estimate the barrier heights was suggested
in Ref.\cite{Cavagna98}. The modification of the dynamical theory allowing
to calculate the barrier heights was suggested in our previous work Ref. 
\cite{Lopatin99}. In the present paper we apply this method to the p-spin
interacting spin-glass model.

The p-spin interacting model  was investigated by the replica \cite{cs},
dynamical \cite{chs,ck}, and Thouless-Anderson-Palmer (TAP) \cite{cs2}
methods . The qualitative picture arising from  these studies is the
following. (i) The transition to the spin-glass state is discontinuous.
Replica and dynamical theories give different results for the transition
temperature, physically the former corresponds to the equilibrium transition
and the latter to the dynamical freezing preempting it. (ii) At all
temperatures lower than the dynamical transition temperature there is an
exponential number of metastable states  \cite{cs2} with 
energies distributed in some interval ($E_{m},E_{CK}$). 
For a typical state with the energy less than $E_{CK}$
the Hessian of fluctuations has strictly positive eigenvalues so such states 
are  stable, for energies larger than $E_{CK}$ some of these eigenvalues become 
negative so a typical state is unstable; for energies lower than $E_{m}$ 
there are no states.
 (iii) Dynamical evolution starting from a random configuration leads to a
non equilibrium regime exhibiting very slow power law relaxation (aging) 
\cite{ck}, in the following we shall refer to it as Cugliandolo-Kurchan (CK)
regime. 

The structure of the phase space described above allows one to understand
better the physics underlying the aging relaxation found in \cite{ck}. The
number of states of a given energy, ${\mathcal N}_{E}$, is exponentially
large, $S_{E}=\frac{1}{N}\ln {\mathcal N}_{E}\sim O(1)$ for all energies
larger than $E_{m}$ ($S_{E_{m}}=0$); furthermore the complexity, $S_{E}$, is
a monotonic function of the energy, $E$. Thus, a set of all locally stable
states is dominated by the states with highest energy, $E_{CK}$, for which
these states are still locally stable and dynamics starting from random
initial conditions is most likely to end up in one of these most abundant
states. A typical state of the largest energy has one eigenvalue which is
very close to zero, so the dynamical evolution leads to a marginally stable
state but it approaches it very slowly, in agreement with the results of
direct dynamical study \cite{ck}. This qualitative argument can be made more
precise. At time scales $t $ \emph{weakly} unstable states with
eigenvalues $\lambda >-1/t \,$are allowed. Because for the p spin model
the maximal energy is directly related to the minimal allowed eigenvalue the
states with $\lambda _{\min }\approx -1/t $ dominate the dynamics at time
scales $t $. These states are, of course, unstable with the average
exponent of the order $\lambda _{\min }$; for the correlation function it
means that 
$$
\partial _{t}\,q(t,t^{\prime })=-\frac{\gamma }{t }q(t,t^{\prime })
$$
with $\gamma \sim 1$. The solution of this equation is a power law
relaxation, consistent with the result of the rigorous study \cite{ck}.
Another conclusion of this reasoning is that the upper bound $E_{CK}$ should
be equal to the asymptotic energy of CK regime as, indeed, was found in \cite
{cs2}. Thus, the dynamics  starting from random initial conditions
(i.e. from a state with $E=0$) leads to the equilibrium states with $E<E_{CK}
$ only in exponentially rare cases. However, as we will show, these
equilibrium states can be reached by the dynamical evolution if the system
starts from a state with a sufficiently low initial energy, $E<E_{c}$. This
will allow us to study the properties of these states in the framework of
the dynamical theory.

Our goal is to find the probability, ${\mathcal P}$, for the system to escape
one of the highest energy TAP states using the instanton method. Generally, $%
{\mathcal P}$ $\sim \exp\, N (-E_{b}/T+S)$ where $E_{b}$ is the energy barrier
corresponding to this transition and $S$ is the entropy (configurational)
contribution. Note the important difference with the usual problem of activated 
transitions, there the contribution from the unique saddle point should be taken 
into account whereas here the number of saddle points is also exponential which 
modifies the result. The dependences  of the energy barrier $E_{b}$ and  action 
$A=-E_{b}/T+S$ on the temperature are  plotted in Fig.1. As one expects, 
the barriers emerge at the dynamical transition temperature. At low temperatures 
the entropy contribution to the action is negligible and the action is 
$A=-E_{b}/T.$

\begin{figure}[here]
\unitlength1.0cm
\begin{center}
\begin{picture}(8,5)
\epsfxsize=8.0 cm
\epsfysize=5.0 cm
\epsfbox{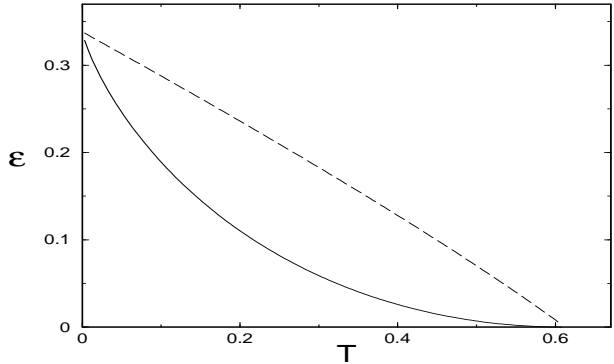}
\end{picture}
\caption{ The effective energy  $\epsilon=-(T\ln {\mathcal P})/N $ controlling
probability of the decay of metastable states (full line) for
$p=3$; the  dashed line shows the energy barrier $E_b.$}
\end{center}
\end{figure}
\vspace{-0.3 cm}

  The idea of our method is the following:
Consider the system that is initially in one of the metastable
states characterized by the energy $E_{in}$, we want to find the probability
to escape this state. Let us consider all saddle point trajectories
(instantons) leading to transient states with energy $E_{tr}>E_{in}$.
Technically we can choose such trajectories inserting an appropriate
delta-function constraint in the dynamical functional. After this process
the system is left to relax. Apparently, if the transient energy, $E_{tr}$,
is high enough, the system will relax to the state which is different from
the initial one, in this case the spin-spin correlation function $\langle
S(t_{i})S(t_{f})\rangle $ is zero. On the contrary, if the energy $E_{tr}$
is just a little above the energy of the initial stable state $E_{in}$, the
system will relax to the same state and the spin-spin correlation function $%
\langle S(t_{i})S(t_{f})\rangle $ will be  equal to the Edwards-Anderson (EA)
order parameter. As we will show, there is a critical value, $E_{c}$, of the
energy $E_{tr}$, such that when $E_{tr}<E_{c}$ the system relaxes back to
the same state while for $E_{tr}>E_{c}$ the system escapes. The difference
between the critical value of the transient energy $E_{c}$ and the energy of
the initial state $E_{in}$ is exactly the barrier energy $E_{b}=E_{tr}-E_{in}$,
and the probability of the instanton transition is determined by the action
of the instanton process corresponding to the transition over this barrier.

We now provide the details of the calculation. Using the functional
formulation of the dynamical theory (see, for example Ref.\cite{fh}) we
write the spin correlation function as a functional integral

\begin{eqnarray}
\langle S(t_{i})&&S(t_{f})\rangle =\int
DS\,\,\,S(t_{i})\,\,\,e^{\int_{t_{i}}^{t_{tr}}{\mathcal L}[S(t)]\,dt} 
\nonumber \\
&&\times \delta ({\mathcal H}(t_{tr})-N E_{tr})\,\,\,e^{\int_{t_{tr}}^{t_{f}}%
{\mathcal L}[S(t)]\,dt}S(t_{f}),  \label{corr}
\end{eqnarray}
where the Lagrangian is

\begin{equation}
{\mathcal L}=\sum_{i}-\hat{S}_{i}^{2}-i\hat{S}_{i}\left( \partial _{t}S_{i}+{%
\frac{{\delta \beta {\mathcal H}}}{{\delta S_{i}}}}\right) +{\frac{1}{2}}{%
\frac{{\delta ^{2}\beta {\mathcal H}}}{{\delta S_{i}^{2}}}},  \label{lagr}
\end{equation}
${\mathcal H}$ is the Hamiltonian of the p-spin interacting model

\begin{equation}
{\mathcal H}=-{\frac{1}{{p!}}}\sum_{i1,i2...ip}J_{i1,i2...ip}\,%
\,S_{i1}S_{i2}...S_{ip}.
\end{equation}
and $J_{i1,i2...ip}$ is a random symmetric tensor normalized by $\langle
J_{i1,i2...ip}^{2}\rangle =p!/2 N^{p-1}$

 The instanton motion from $t_{i}$ to $t_{tr}$ can be transformed into the usual
causal motion with the help of the transformation\cite{Lopatin99}
\begin{eqnarray}
S_{1}(-t) &=&S(t),  \label{tr1} \\
i\hat{S}_{1}(-t) &=&-\partial_t S(t)+i\hat{S}(t),  \label{tr2}
\end{eqnarray}
where $t_{tr}=0$ was taken to make formulas more compact.
In order to distinguish this time interval from the relaxational motion at $
t>t_{tr}=0$ we denote $S$ field 
\begin{equation} 
S_{2}(t) =S(t), \;\;\; \hat{S}_{2}(t)=\hat{S}(t)
\end{equation}
for $t>0.$
In the new notations the correlation function (\ref{corr}) becomes

\begin{eqnarray}
&&\langle S(t_{i})S(t_{f})\rangle =e^{-\beta \Delta{\cal  E}}\int
DS_{1}\,\,DS_{2}\,\,e^{\int_{0}^{-t_{i}}{\mathcal L}[S_{1}(t)]\,dt}  \nonumber
\\
&&\times e^{\int_{0}^{t_{f}}{\mathcal L}[S_{2}(t)]\,dt}\delta ({\mathcal H}%
(0)-N E_{tr})S_{1}(-t_{i})S_{2}(t_{f}),  \label{tworep}
\end{eqnarray}
where $\Delta {\cal E}={\mathcal H}(0)-{\mathcal H}(t_{i}).$

We see that Eq.(\ref{tworep}) corresponds to the two-real-replica problem
with a fixed initial energy. The Green functions become matrices in the
replica space
$$
D_{\alpha,\beta}=\langle S_\alpha S_\beta\rangle,\,\,\,\,G_{\alpha,\beta}
=\langle S_\alpha\, i\hat
S_\beta\rangle ,\,\,\,\,\hat D_{\alpha,\beta}=\langle i\hat 
S_\alpha \,i\hat S_\beta\rangle.
$$
The advantage of this representation is that since Eq.(\ref{tworep})
describes the normal (relaxational) motion, the response functions 
$G_{\alpha,\beta}$ are casual and the anomalous Green functions 
$\hat{D}_{\alpha,\beta}\equiv 0$. Moreover,
according to Eq.(\ref{tworep}) the replicas are independent from each other,
therefore 
\begin{equation}
G_{1,2}(t_{1},t_{2})=G_{2,1}(t_{1},t_{2})=0.
\end{equation}
Also we will be interested only in the replica-symmetric processes, for
which one can write 
\begin{eqnarray}
D_{1,1}(t_{1},t_{2}) &=&D_{2,2}(t_{1},t_{2})\equiv D(t_{1},t_{2}), \\
D_{1,2}(t_{1},t_{2}) &=&D_{2,1}(t_{1},t_{2})\equiv D^{\prime }(t_{1},t_{2}),
\\
G_{1,1}(t_{1},t_{2}) &=&G_{2,2}(t_{1},t_{2})\equiv G(t_{1},t_{2}).
\end{eqnarray}
Representing the delta-function in (\ref{tworep}) as 
\begin{equation}
\delta ({\cal H}(0)-N E_{tr})=\int d\phi \,\,\,e^{\,\,i\phi 
({\cal H}(0)-N\,E_{tr})\beta },
\label{constraint}
\end{equation}
we perform averaging over the disorder.  Then, making the saddle point
approximation we get the equations for the correlation functions:

\begin{eqnarray}
[{\partial}_{t1}+1-p\beta E(t_1)]G(t_1,t_2)-\delta(t_1-t_2)  \nonumber \\
-\mu(p-1)\int_{t1}^{t2} D^{p-2}(t_1,t) G(t_1,t)G(t,t_2)\,dt=0,  \label{g}
\end{eqnarray}

\begin{eqnarray}
&[{\partial}_{t1}+1-p\beta E(t_1)]D(t_1,t_2) +p\beta E(0) D^{p-1}(t_1,0)
D(0,t_2)  \nonumber \\
&-\mu(p-1)\int_{0}^{t1}D^{p-2}(t_1,t) G(t_1,t)D(t,t_2)\,dt  \nonumber
\end{eqnarray}
\vspace{-0.9 cm}
\begin{equation}
-\mu\int_{0}^{t2}D^{p-1}(t_1,t) G(t_2,t)\,dt=0,  \label{d}
\end{equation}
\begin{eqnarray}
& [{\partial}_{t1}+1-p\beta E(t_1)]\,D^\prime(t_1,t_2) +p\beta E(0)
D^{p-1}(t_1,0) D^\prime(0,t_2)  \nonumber \\
&-\mu(p-1)\int_{0}^{t1}D^{p-2}(t_1,t) G(t_1,t)D^\prime(t,t_2)\,dt  \nonumber
\end{eqnarray}
\vspace{-0.9 cm}
\begin{equation}
-\mu\int^{t2}_{0}{D^\prime}^{p-1}(t_1,t) G(t_2,t)\,dt=0 ,  \label{dp}
\end{equation}
where the energy $E$ is
\begin{equation}
E(t) =-{\beta p \over 2}\int_{0}^{t}dt^{\prime
}D^{p-1}(t,t^{\prime })G(t,t^{\prime })+E(0)D^{p}(t,0),  \label{e} 
\end{equation}
$E(0)=E_{tr}$ and $\mu =p\beta ^{2}/2.$ The value of $\phi $ 
in Eq.(\ref{constraint})
at the saddle point is related to the initial energy $E(0)$ by $i\phi
=E(0)/\beta .$  Note that Eqs.(\ref{g},\ref{d},\ref{e})
describe the usual one-replica dynamics, because they do not contain the
correlation function $D^{\prime }.$ Therefore, in order to find the most
interesting correlation function $D^{\prime }$ we need first to find $G$ and 
$D$ solving Eqs.(\ref{g},\ref{d},\ref{e}) and then find $D^{\prime }$ from
Eq.(\ref{dp}).

Thus, we begin with the analysis of the one-replica equations (\ref{g},\ref
{d},\ref{e}).  When $E(0)=0$ the terms emerging from the energy constraint 
in Eqs.(\ref{d},\ref{e}) disappear  and the asymptotics of the correlation
functions should have the aging form found in Ref.\cite{ck} 
$$
D(t_{1},t_{2}) \sim \left( {\frac{{t_{2}}}{{t_{1}}}}\right) ^{\gamma},\,
G(t_{1},t_{2}) \sim \left( {\frac{{t_{2}}}{{t_{1}}}}\right) ^{\gamma-1},
\,\,t_{1}>t_{2}\gg 1,
$$
where $\gamma $ is a constant depending on temperature. Numerical analysis
of Eqs.(\ref{g},\ref{d},\ref{e}) shows that this aging regime is still
realized when $E(0)$ is higher than some negative critical value $E_{c},$
which we will find analytically later.

For $E(0)<E_{c}$ the numerical solution shows that the system relaxes to a
stable state with correlation functions satisfying FDT at large times $%
(t_{1},t_{2}\gg 1).$ In this case the limiting value of the energy, $E_{in}$%
, and the Edwards-Anderson order parameter, $q$, characterizing this state
can be found analytically: At large times $t_{1},t_{2}\gg 1$ the Green
functions depend only on the time difference $t_{1}-t_{2}=\tau $ and FDT
holds $\partial _{\tau }D(\tau )=-G(\tau ).$ Using these facts one can
integrate Eq.(\ref{g}) over $\tau $ obtaining 
\begin{equation}
-p\beta \ E_{in}=\mu (1-q^{p-1})+{\frac{q}{{1-q}}},  \label{1}
\end{equation}
where $q=D(t+\tau ,t),\,\,\,\,t,\tau \gg 1.$ 
Considering Eq.(\ref{d}) on times $t_{1}\gg 1,\,\,t_{2}=0$ we get 
\begin{equation}
-p\beta \ E_{in}=\mu (1-q^{p-1})-p\,\beta \,q_{1}^{p-2}E_{tr}-1,  \label{2}
\end{equation}
where $q_{1}$ is the overlap between the initial and final states 
$
q_{1}=D(t,0),\,\,\,\,t\gg 1.$
The equation  closing  the system (\ref{1},\ref{2}) follows form Eq.(\ref{e}) 
\begin{equation}
-p\beta \ E_{in}=\mu (1-q^{p})-p\,\beta \,q_{1}^{p}E_{tr}.  \label{3}
\end{equation}
 Combining Eqs.(\ref{1},\ref{2},\ref{3}) and introducing  a new variable 
$z=\beta (1-q)\,\,q^{(p-2)/2}$ we get the equation
\begin{equation}
{\frac{p}{2}}\,z^{2}+\left( p|E_{tr}|\,z\right) ^{-{\frac{2}{{p-2}}}}=1,
\label{z}
\end{equation}
which contains only one parameter $E_{tr}$. Therefore, all the temperature
dependence of $q$ (at fixed $E_{tr}$) comes from the equation $z=\beta
(1-q)\,\,q^{(p-2)/2}$, this  agrees with TAP approach of Ref.\cite{cs2}. 
One can show that Eq.(\ref{z}) has solutions only when $E_{tr}<E_{c}$, with 
\begin{equation}
E_{c}=-(2p)^{-1/2}(p-2)^{(2-p)/2}(p-1)^{(p-1)/2},
\end{equation}
which is related to the dynamical transition temperature $T_{c}$ by $%
E_{c}=-1/2T_{c}.$ For $E_{tr}<E_{c}$ Eq.(\ref{z}) has two solutions but
only one of these solutions is physical, it lies in the interval $0<z<z_{c}$
with $z_{c}=\sqrt{2/(p-1)p}.$

Eqs.(\ref{1},\ref{2},\ref{3}) can be solved analytically for one special
value of the energy $E_{tr}=-\beta /2.$ In this case $E=E_{tr},\,\,q_{1}=q$
and Eqs.(\ref{1},\ref{2},\ref{3}) are reduced to a single equation 
\begin{equation}
\mu \,\,q^{p-2}(1-q)=1.  \label{spec}
\end{equation}
This special value of the energy has a simple physical meaning: The system 
remains at the same state and it is at equilibrium at all times during the 
process.

To summarize the results of one-replica dynamics: When $E_{tr}>E_{c}$ the
system relaxes to the aging CK state, when $E_{tr}<E_{c}$ the system relaxes
to one of the locally stable states. There is a special value $E_{tr}=-\beta /2$ 
such that the system is at the equilibrium for all times.

Now turn to the behavior of the correlation function $D^{\prime }.$
Numerical solution of Eq.(\ref{dp}) gives the following asymptotics of the
correlation function $D^{\prime }:$ 
\begin{eqnarray}
E_{tr} &<&E_{c}:\,\,\,\,D^{\prime }(t_{1},t_{2})\to
q_{EA},\;\;\;t_{1},t_{2}\to \infty ,  \label{adp2} \\
E_{tr} &>&E_{c}:\,\,\,\,D^{\prime }(t_{1},t_{2})\to 0,\;\;\;\; \;t_{1},t_{2}\to
\infty.  \label{adp1}
\end{eqnarray}

It means that for energies $E_{tr}<E_{c}$ the replicas relax to the same
state, while for $E_{tr}>E_{c}$ to different ones. For the original problem
it means that when $E_{tr}<E_{c}$ the system remains in the same TAP state
after the instanton process, while for $E_{tr}>E_{c}$ the system escapes
the original state. The result (\ref{adp2}) can also be obtained
analytically by considering Eq.(\ref{dp}) and assuming that FDT holds at
large times. Thus  we conclude that $E_{b}=E_{c}-E_{in}$ is the energy
of the barriers between the highest energy TAP states. The height of the
barrier $E_{b}$ as a function of temperature is plotted on Fig.1.

The probability to escape a metastable state is $e^{N A}$ where the action $A$
is

\begin{equation}
A={1\over N}\ln \int DS\,\,\,e^{\int_{t_{i}}^{0}{\mathcal L}(S)}.  \label{a}
\end{equation}
The differentiation of (\ref{a}) with respect to $\beta $ gives

$$
{\frac{{dA}}{{d\beta }}}={\frac{1}{{N (p-1)!}}}\int_{t_{i}}^{t_{ins}}\left%
\langle \sum J_{i1,i2...ip}\,\,i\hat{S}_{i1}S_{i2}...S_{ip}\right\rangle
_{S,J}.
$$
Using the transformation law (\ref{tr1},\ref{tr2}) we get 
\begin{equation}
{\frac{{dA}}{{d\beta }}}={{-{\mathcal H}(0)+{\mathcal H}(t_{i})}\over N}
=-E_{tr}+E_{in}.  \label{ae}
\end{equation}
To integrate (\ref{ae}) it is convenient to use the result for the free
energy of TAP states of Ref.\cite{cs2} 
\begin{eqnarray}
f_{in} &=&-q^{{\frac{p}{2}}}\left( {\frac{1}{{pz}}}+{\frac{1}{2}}%
z(p-1)\right) -{\frac{T}{2}}\ln (1-q)  \nonumber \\
&&-{\frac{\beta }{4}}\left( 1+(p-1)q^{p}-pq^{p-1}\right).  \label{f}
\end{eqnarray}
Note that this expression does not contain  the complexity contribution. 
Using
Eqs.(\ref{z},\ref{f}) one can check that 
\begin{equation}
E_{in}={\frac{d}{{d\beta }}}\,\,\beta f_{in}.
\end{equation}
Therefore for the action we have 
\begin{equation}
A=\beta f_{in}-E_{tr}\beta -\Gamma (E_{tr}),  \label{A/N}
\end{equation}
where $\Gamma $ is an unknown function which depends only on $E_{tr}.$
 This function can be found from the following arguments. As was
shown above there is a special value of $E_{tr}=-\beta /2$ for which the
system is at equilibrium for all times. It means that actually there is no
instanton motion, therefore the action of this process is zero. Imposing
this requirement and using Eq.(\ref{spec}) we find $\Gamma $, inserting it
in (\ref{A/N}) we finally get  
\begin{equation}
-A=E_{tr}^{2}+E_{tr}\beta -[\beta f_{in}-g(z)],  \label{act}
\end{equation}
where 
\begin{equation}
g(z)={\frac{1}{2}}\left( {\frac{{2-p}}{p}}-\ln {\frac{{pz^{2}}}{2}}+{\frac{{%
p-1}}{2}}z^{2}-{\frac{2}{{p^{2}z^{2}}}}\right) ,
\end{equation}
and $z$ is determined by Eq.(\ref{z}). The function $g(z)$ turns out to be
the same as the complexity of the free energy minima found in Ref.\cite{cs2}
In order to get the action which determines the probability to
escape a metastable state one should take $E_{tr}=E_{c}$ in (\ref{act}).
The result is plotted in Fig.1.

The result (\ref{act}) can be presented in a very simple form if we
introduce the ''full'' free energy including complexity following 
Ref.\cite{cs2} 
\begin{equation}
\bar{f}(T)=f_{in}-Tg-T(1+\ln (2\pi ))/2,
\end{equation}
where the third term corresponds to  the conventional \cite{cs2} normalization 
of the trace over spins. We define the 'free energy' of a random 
state $f_{R}$ with a given energy $E$ as
\begin{equation}
f_{R}(E)=E-TS_{R},
\end{equation}
where $S_{R}$ is the 'entropy' of the random state defined by 
\begin{equation}
\left\langle \delta (NE-{\mathcal H}(S)\right\rangle _{J}=e^{N\,S_{R}}
\end{equation}
The straightforward calculation gives $S_{R}(E)=(\ln 2\pi +1)/2-E^{2}.$
Then the action $A$ becomes 
\begin{equation}
-A=[f_{R}(E_{tr},T)-\bar{f}(T)]/T.
\end{equation}
We see that with these  definitions of the free energies 
the action has a form corresponding to the Gibbs rule.

In conclusion, we found the energy barriers between metastable states with
energies $E \approx E_{CK}$, $E < E_{CK}$ using the instanton method. 
Mapping of the
instanton motion into the usual motion going back in time allows to reduce
the problem to the causal two-real-replica dynamics. The probability to
escape a metastable state is determined by the action of the instanton
process, this action is expressed through the ''full'' free energy of the
system that includes complexity. The method developed in the paper can be
applied to study the free energy structure of other spin-glasses. The
drawback of this method is that it does not allow one to find all saddle
point trajectories, instead it finds only a typical one, corresponding to a
typical barrier; in particular, in the p-spin model we were able to study
only the transitions between the most abundant states with $E\approx E_{CK}.$
 Ideally one wants to know  the barrier (or distribution of
barriers) separating two states with given energies; this is a much more
difficult problem which remains unresolved.

\end{document}